\documentclass[12pt,preprint]{aastex}
\begin{document}

\newcommand{\kms}{\mbox{km~s$^{-1}$}}
\newcommand{\s}{\mbox{$''$}}
\newcommand{\mloss}{\mbox{$\dot{M}$}}
\newcommand{\my}{\mbox{$M_{\odot}$~yr$^{-1}$}}
\newcommand{\ls}{\mbox{$L_{\odot}$}}
\newcommand{\ms}{\mbox{$M_{\odot}$}}
\newcommand\mdot{$\dot{M}$}
\def\arcdeg{\hbox{$^\circ$}}
\def\farcs{\hbox{$.\!\!^{\prime\prime}$}}
\def\gtrsim{\mathrel{\hbox{\rlap{\hbox{\lower4pt\hbox{$\sim$}}}\hbox{$>$}}}}
\def\lesssim{\mathrel{\hbox{\rlap{\hbox{\lower4pt\hbox{$\sim$}}}\hbox{$<$}}}}

\title{A Massive Bipolar Outflow and a Dusty Torus with Large Grains in
the Pre-Planetary Nebula IRAS\,22036+5306}
\author{R. Sahai\altaffilmark{1}, K. Young\altaffilmark{2}, N.A. 
Patel\altaffilmark{2},  
C. S\'anchez Contreras\altaffilmark{3}, M. Morris\altaffilmark{4}
}

\altaffiltext{1}{Jet Propulsion Laboratory, MS 183-900, California
Institute of Technology, Pasadena, CA 91109}

\altaffiltext{2}{Harvard-Smithsonian Center for Astrophysics, Cambridge}

\altaffiltext{3}{California Institute of Technology, MS\,105-24, 
Pasadena, CA 91125, {\it current address}: Dpto. de Astrofisica Molecular e
Infraroja, IEM-CSIC, Serrano 121, 28006 Madrid, Spain}

\altaffiltext{4}{Division of Astronomy, Department of Physics and
Astrophysics, UCLA, Los Angeles, CA 90095}

\email{raghvendra.sahai@jpl.nasa.gov}

\begin{abstract} We report high angular-resolution ($\sim$1$''$) CO J=3--2
interferometric mapping, using the Submillimeter Array (SMA), of
IRAS\,22036+5306 (I\,22036), a bipolar pre-planetary nebula (PPN) with
knotty jets discovered in our HST SNAPshot survey of young PPNs. In
addition, we have obtained supporting lower-resolution ($\sim$10$''$) CO
and $^{13}$CO J=1--0 observations with the Owens Valley Radio Observatory
(OVRO) interferometer, as well as optical long-slit echelle spectra at the
Palomar Observatory. The CO J=3--2 observations show the presence of a very
fast ($\sim$220\,\kms), highly collimated, massive (0.03\ms) bipolar
outflow with a very large scalar momentum (about $10^{39}$ g cm s$^{-1}$),
and the characteristic spatio-kinematic structure of bow-shocks at the tips
of this outflow. The H$\alpha$ line shows an absorption feature
blue-shifted from the systemic velocity by $\sim$100\,\kms, which  
most likely arises in neutral interface material between the fast outflow
and the dense walls of the bipolar lobes at low latitudes. Since the
expansion age of the outflow as determined from our data is only about 25
years, much smaller than the
time required by radiation pressure to accelerate the observed bipolar
outflow to its current speed ($\gtrsim10^5$ yr), the fast outflow in
I\,22036, as in most PPNs, cannot be driven by radiation pressure. The
total molecular mass (0.065\,\ms) is much less than that derived from a
previous detailed model of the near to far-infrared SED of this object
(showing the presence of a large, cool dust shell of total mass 4.7\ms),
most likely due to the interferometric observations resolving out the CO
flux from the dust shell due to its size. We find an unresolved source of
submillimeter (and millimeter-wave) continuum emission in I\,22036,
implying a very substantial mass (0.02-0.04\,\ms) of large
(radius$\gtrsim$1\,mm), cold ($\lesssim$50\,K) dust grains associated with
I\,22036's toroidal waist. We also find that the $^{13}$C/$^{12}$C ratio in
I\,22036 is very high (0.16), close to the maximum value achieved in
equilibrium CNO-nucleosynthesis (0.33). The combination of the high
circumstellar mass (i.e., in the extended dust shell and the torus) and the
high $^{13}$C/$^{12}$C ratio in I\,22036 provides strong support for this
object having evolved from a massive ($\gtrsim$4\,\ms) progenitor in which
hot-bottom-burning has occurred.

\end{abstract}
\keywords{circumstellar matter -- planetary nebulae: individual (IRAS
22036+5306) -- reflection nebulae -- stars: AGB and post-AGB -- stars: mass
loss -- stars: winds, outflows
}
\section{INTRODUCTION}

Pre-Planetary nebulae (PPNs) - objects in transition between the AGB and
planetary nebula (PN) evolutionary phases - hold the key to one of the
most vexing and long-standing problems in our understanding of these very
late stages of evolution for low and intermediate mass stars. Current
observational evidence shows that during the PPN phase, the dense,
isotropic mass-loss which generally marks the late AGB phase changes to
high-velocity, bipolar or multipolar mass-loss. The emergence of fast
collimated
outflows or jets has been hypothesized as the primary mechanism for this
dramatic change in the geometry and dynamics of the mass-loss (Sahai \&
Trauger 1998). However, the physical mechanism for producing the fast
outflows remains unknown. (Sub)millimeter-wave interferometric
observations offer one of the best probes of the dynamics and energetics
of the shock acceleration process, which transfers a substantial amount
of directed momentum to large parts of the dense AGB wind.

IRAS\,22036+5306 (I\,22036) is a bipolar PPN with knotty jets discovered
in our HST SNAPshot survey of young PPNs, with an estimated distance and
luminosity of 2\,kpc and 2300\ls, respectively (Sahai et al. 2003
[SZSCM03]). 
Our HST imaging suggested that I\,22036 is a key object for understanding
how jets can sculpt bipolar lobes in a progenitor AGB star wind, since
its jets and their working surfaces appear to be simultaneously observed.
In
this paper (the second one by Sahai and collaborators on this important
PPN) we report CO J=3--2 observations of the PPN with the
Submillimeter Array (SMA), which resolve a massive high-velocity bipolar
outflow aligned with the optical nebulosity and clearly pinpoint the
location and nature of the bow-shocks at the tips of this outflow.
We also report supporting lower-resolution
($\sim$10$''$) CO and $^{13}$CO J=1--0 observations with the Owens Valley
Radio Observatory (OVRO) interferometer, as well as optical long-slit
echelle spectra at the Palomar Observatory.

The rest of the paper is organised as follows: in \S\,2 we describe our 
observational techniques and data-reduction, in \S\,3 we
describe our observational results, in \S\,4 we discuss the kinematics, age
and mass of the bipolar outflow, in \S\,5 we derive the total molecular
mass and compare it to the mass of dust in 
the circumstellar environment, in \S\,6 we derive and discuss the
implications of the $^{13}$C/$^{12}$C isotope ratio in I\,22036, in \S\,7
we discuss the
formation of the nebula, and in \S\,8 we summarize our main conclusions.

\section{OBSERVATIONS}

The CO J=3--2 observations were obtained with the SMA, using two tracks,
on July 20 (UT) and Aug. 29 (UT), 2004, with the array in the compact and
extended
configurations, respectively. The J2000 coordinates used for I22036 were
RA = 22$^{h}$05$^{m}$30.28$^{s}$, Dec = +53$\arcdeg$21$'$33.0$''$. The
receivers were tuned to center the 345.796 GHz rest frequency of the
CO(3--2) line in the upper sideband (USB), after correcting for the source's
-45 kms$^{-1}$ LSR velocity. On the night of July 20th, the weather was
excellent, with 0.7 mm of precipitable water vapor. The antennas' DSB
system temperatures ranged from $\sim 200$K near the zenith to $\sim
350$K at an elevation of 30$\arcdeg$. Unfortunately one of the eight
antennas had extremely high phase noise on that night, and its data was
unusable. On the night of Aug. 29, all eight antennas were working well,
but the precipitable water vapor varied between 1.0 to 1.6 mm, causing
the system temperatures to range from $\sim 250$K to $\sim 500$K. For
both tracks the correlator was configured to have the highest available
uniform resolution, 406.25 kHz channel (0.35 kms$^{-1}$), over a total
bandwidth of 2 GHz. Observations of Saturn and Uranus were obtained for
bandpass calibration, and Titan and Uranus were used for flux calibration.
The
quasars BL Lac and 2013+370 were observed every 1/2 hour for use as
complex gain calibrators. A total of 12 hours of on--source data was
taken, split nearly evenly between the two tracks.

The data were calibrated using the MIR package, and mapping was done with
AIPS. Visibility
weights were derived from receiver Tsys measurements.   After calibration,
the data was exported to AIPS for mapping. Two spectral lines, CO(3-2)
and an unexpected second line, were detected in the USB, but
no lines were seen in the LSB. A continuum level was fitted to
the regions of the USB data which were free of line emission, and
subtracted from the visibility data. Two data cubes were then generated.
When producing the first cube, we 
included data from both tracks with the aim of generating maps of the
emission integrated over large velocity intervals with the smallest
synthesized beam.  For the second, we discarded the more noisy data from the
extended track, with the aim of generating high-quality emission maps with
much narrower velocity bins. The final data cubes had 4872 channels after
the removal of overlapping portions of the spectra lying in the rolloff
regions of the hybrid correlator's analog filters. In the first (second)
data cube, the clean beam had a
FWHM$=1\farcs12 \times 1\farcs00$ ($1\farcs45 \times 1\farcs29$) at a
position angle (PA) of $88^{\circ}.6$ ($-84^{\circ}.7$). A continuum map
was made using the lower sideband (LSB) data, and uniform weighting of
Tsys-weighted visibilities. The central 250 MHz of the LSB IF coverage was
rejected, in order to guard
against contamination from the strong CO emission in the USB. A 180
k$\lambda$ taper
was applied to the visibility data when maps were produced. Uniform
weighting was used.

Interferometric observations of the CO and $^{13}$CO J=1--0 line emission
were obtained at OVRO on 2002 September 21 as part of a snapshot survey
of pre-planetary nebulae. General details on the observational setup are
given in  S\'anchez Contreras \& Sahai (2004); data reduction and
calibration steps are similar to those described in Sahai et al. (2005).
The clean beam had a FWHM of $9\farcs8
\times 5\farcs9$ (PA $87^{\circ}$) for CO and $10\farcs2 \times 6\farcs7$
(PA $82^{\circ}$) for $^{13}$CO. Scale conversion factors were 1.6 and
1.5 K/(Jy beam$^{-1}$) for CO and $^{13}$CO.

Since I\,22036 displays a prominent H$\alpha$ emission line, and the
presence of a fast neutral outflow can produce a self-absorption
signature in this line (e.g., S\'anchez Contreras \& Sahai 2001), we also
obtained a high-resolution
($\Delta\lambda$=0.4\AA) spectrum of I\,22036 in order to resolve the
velocity structure of the line, using the Palomar 60-inch spectrograph in
echelle mode. In this mode, the spectrograph provided a total wavelength
coverage of $\sim$3600-9000\AA, with a resolution of about 19000. The
echelle grating had 52.65 lines/mm and a blaze angle of 63$\arcdeg$26$'$.
Cross-dispersion was provided by two fused quartz prisms with a
60$\arcdeg$ apex angle. The slit was 7\farcs4 long, and had a width of
1\farcs4. The spectra were recorded on a thinned, lumogen coated TI
$800\times800$ chip. A Thorium-Argon lamp was used for wavelength
calibration, and the standard star HR7596 was used for flux calibration.
The data were obtained during the nights of 2002, August 4 and 5, and
reduced and calibrated using standard IRAF techniques.

\section{RESULTS}
\subsection{Spectral Profiles}

\begin{figure}
\epsscale{.70}
\plotone{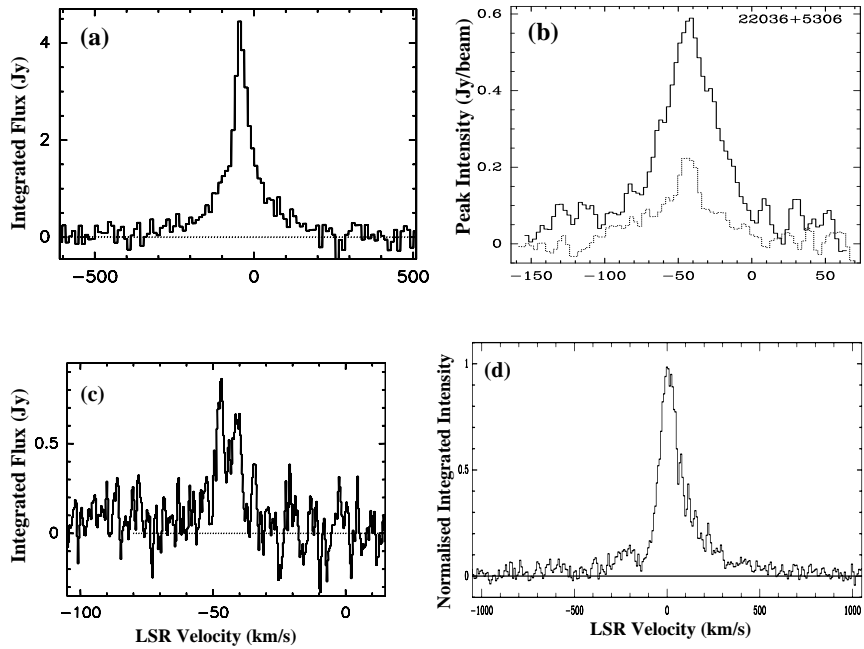}
\caption{The spectra of the (a) CO J=3--2 line integrated over the entire
emission region, (b) CO and $^{13}$CO1--0 peak intensity in the
unresolved emission source for each line, 
(c) the SO 9(8)-8(7) line integrated within a
$2{''}\times$2${''}$ area centered on the SO
peak (5-channel Hanning smoothing has been applied), and
(d) the H$\alpha$ line (flux in erg s$^{-1}$cm$^{-2}$\AA$^{-1}$ 
normalised to a peak value of unity), in IRAS\,22036+5306}. 
\vskip -0.3in
\label{co32spec}
\end{figure}

The (spatially) averaged CO J=3--2 line
profile shows a strong central core and weaker extended wings
(Fig.\,\ref{co32spec}a). The wings cover a total velocity extent (FWZI) of
$\sim$450\,\kms~and are most likely associated with a fast
post-AGB outflow. The large width
(FWHM$\approx$70\,\kms) of the central core of the CO J=3--2 emission
indicates that much of the core emission is also associated with the fast
outflow.

Like the J=3--2 line profile, the CO and $^{13}$CO J=1--0 line profiles
obtained also show wide wings ($\gtrsim$100\,\kms) but do not cover as
wide a velocity range (Fig.\,\ref{co32spec}b). The observed extent of the
CO J=1--0 line wings is limited by the spectrometer bandwidth
($\approx$200\,\kms), whereas that of the $^{13}$CO line wings is
sensitivity limited. The J=1--0 line profile has a triangular shape with
a total velocity extent (FWZI) of greater than 100\,\kms, with no obvious
central core. It is likely that this J=1--0 emission is associated with
the material seen in the central core of the J=3--2 line. The $^{13}$CO
J=1--0 line shows a narrow central core, with a width of about 20\,\kms,
and weaker wings extending over a similar velocity range as the CO J=1--0
line. Since the typical expansion velocities of AGB circumstellar
envelopes (CSEs) are 10--15\,\kms, the narrow central core seen in the
$^{13}$CO J=1--0 line may be associated with the remnant progenitor AGB
envelope in I\,22036, or slowly expanding gas in the dense waist.

\begin{figure}
\epsscale{0.85}
\plotone{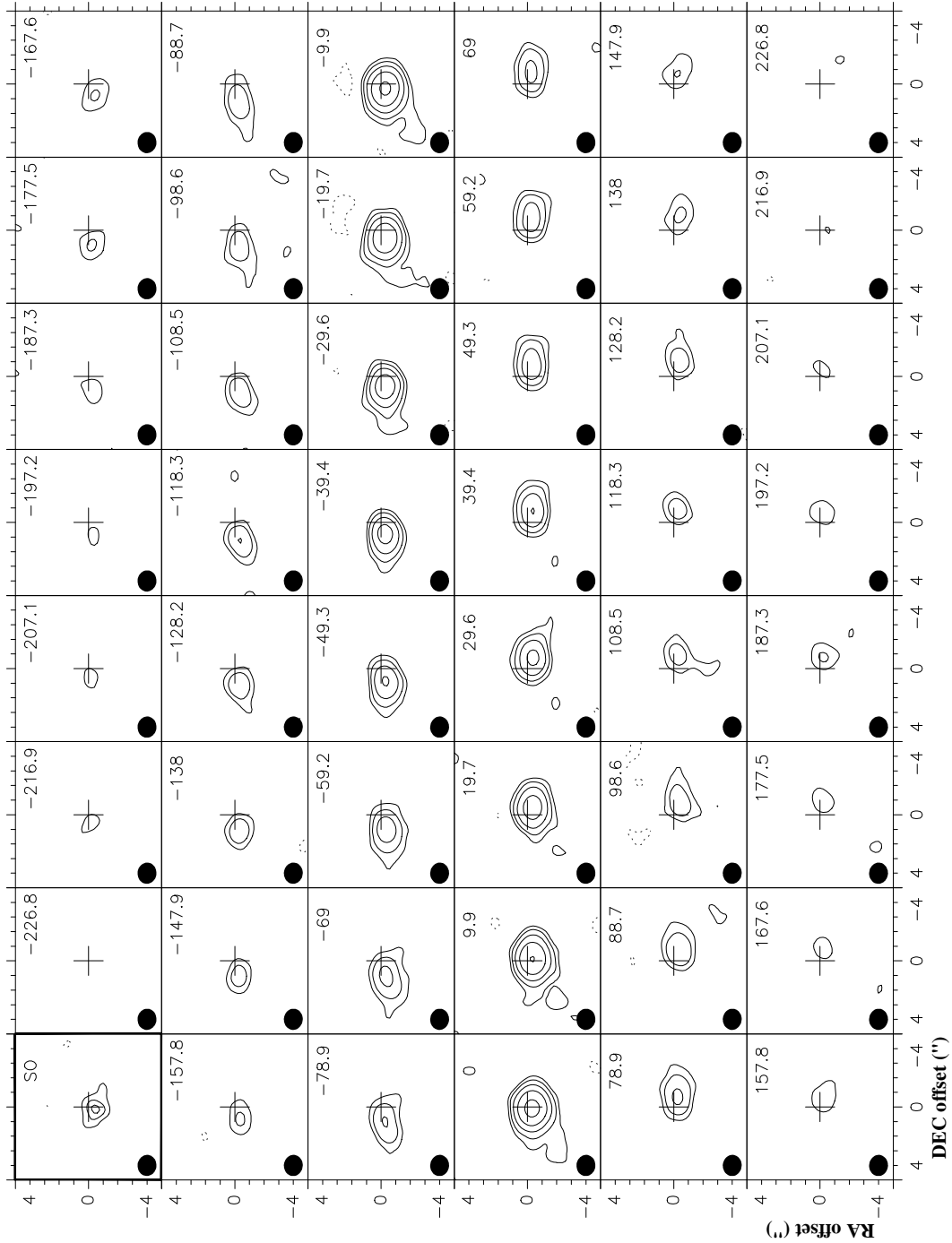}
\caption{Maps of the CO J=3--2 and SO 9(8)-8(7) emission in I\,22036
obtained with the SMA. For SO, the emission is quite weak and only the
(velocity)integrated intensity is shown (top left corner). The
contours on the SO map are -30, 30, 60 and 90\% of the peak flux
of 0.54 Jy/beam. For the CO J=3--2 line, the remaining panels show emission
maps as a function of velocity. Each panel covers a velocity bin
of width 9.86\,\kms, centered at the velocity offset (as measured from the
systemic velocity $V_{LSR} = -45$\,\kms) shown in the top
right corner. The
contour levels are -5, 5, 10, .. 95\% of the peak flux in the central
channel which is 2.2 Jy/beam. The center of each panel (marked by a
cross) is located at RA = 22$^{h}$05$^{m}$30.28$^{s}$, Dec =
+53$\arcdeg$21$'$33.0$''$. The 1$\sigma$ noise is about 38 mJy/beam
}
\label{chanmap32}
\end{figure}

In addition to the CO J=3--2 line, we detected a line at a rest frequency
of 346.53 GHz, most likely due to the SO N=8-7, J=9-8 transition with a
frequency of 346.5284810 GHz (JPL catalog). The observed SO line is much
narrower than the CO line, with a full width of 21\,\kms, and an
integrated flux of 20\,Jy\,\kms (Fig.\,\ref{co32spec}c). The SO emission
appears to be slightly
elongated and extended over $\sim 2''$, with a PA of about $-30\arcdeg$
(Fig.\,\ref{chanmap32}, {\it top-left corner panel}). Within
the relatively low signal/noise, the red and blue wings of the line are
co-located, indicating that the SO emission does not arise in a bipolar
outflow. Given its similar velocity-width compared to the narrow core of
the $^{13}$CO 1--0 line, the SO emission is most likely associated with the
same
material as the former.

The H$\alpha$ line shows a profile with a central peak, very wide wings,
and an absorption dip in the blue wing of the line, centered at $V_{LSR}$
of about $-145$\,\kms~(Fig.\,\ref{co32spec}d). This peculiar H$\alpha$
line shape is seen in many PPNs and young, compact PNs (van de Steene,
Wood, \& van Hoof 2000, S\'anchez Contreras \& Sahai 2001, Sahai \&
S\'anchez Contreras 2004, Sahai et al. 2005). The blue and red wings can be
seen extending to velocities of
roughly $-400$ and 500\,\kms, respectively. The H$\alpha$ line profile %
(e.g. IRAS\,19024+0044, Hen\,3-1475) is most simply interpreted in terms
of the following model -- the (spectrally) broad H$\alpha$ emission
arises in a compact, spatially-unresolved central source\footnote{the
nature of this source is unknown, although possible mechanisms include
emission from a very high velocity outflow, Raman scattering and/or
Keplerian rotation in a dense disk}; this emission and stellar continuum
is scattered by dust in the extended, expanding nebula. The blue-shifted
absorption arises due to a largely atomic outflow within the foreground
lobe, which absorbs the scattered photons. The scattering is expected to
produce an overall red-shift of the H$\alpha$ line peak relative to the
systemic radial velocity, as observed -- the peak of the H$\alpha$ line
is located at $V_{LSR}\approx\,0$\,\kms, red-shifted from the systemic
velocity, $V_{LSR} = -45$\,\kms. Such a model has been successfully used
to quantitatively explain the spatio-kinematic distribution of similar
H$\alpha$ emission profiles observed with STIS/HST in the PPN Hen 3-1475
(S\'anchez Contreras \& Sahai 2001).

\subsection{Velocity-Channel Maps}

We show the spatial distribution of the CO J=3--2 emission as a function
of velocity in Fig.\,\ref{chanmap32}. The centroid of the emission is seen
to shift steadily from west to east in the panels as the average velocity 
decreases from 217 to -217\,\kms, indicating a strong velocity gradient
along the major axis of the nebula. A plot of the central
$\pm$15\kms~range with narrower velocity bins of width 2.47\,\kms
(Fig.\,\ref{chancen32}) also shows roughly a East-West velocity gradient.
We cannot
separate a kinematically distinct component in the J=3--2 emission in the
velocity range covered by the central core of the $^{13}$CO J=1--0 line.
The overall velocity gradient can be seen more easily in a 
plot of the blue and red wing emission of the CO J=3--2 line, overlaid on
the optical image of I\,22036: there is a
large separation between the red and blue-shifted emission, implying the
presence of a 
bipolar outflow (Fig.\,\ref{co32wings-hst}) along the long axis of
I\,22036. In order to register the HST and SMA images, we first rotated the
latter to the orientation of the former. For the final fine translational
registration, we assumed that peak ``C" SZSCM03 coincides with the peak of
the integrated CO J=3--2 intensity (shown in Fig.\,\ref{co32-hstbw}) since
the absolute astrometry of the HST images, relative to that of the SMA
images, cannot be determined to a sufficently high accuracy (i.e., better
than few$\times 0\farcs1$). 
The outflow is tilted such that the W-lobe (E-lobe) is on the near
(far) side -- this tilt is consistent with that inferred from the
difference in brightness between the two optical lobes and the morphology
of the waist (SZSCM03). We have not shown the
corresponding integrated intensity or velocity channels maps for the OVRO
observations, because these do not resolve the CO J=1--0
emission due to the large beamwidth.

The map of the emission integrated over all velocities 
Fig.\,\ref{co32-hstbw} shows that the 
the emission lies in an elongated region, oriented along
the major axis of the optical nebula.
A central,
unresolved, continuum source (at $\lambda$=0.88\,mm) with a flux of
0.29$\pm$0.04\,Jy is also seen; the error is dominated by calibration
uncertainties. The peak of the continuum source appears to be offset from
the line intensity peak by $\sim$0\farcs23. 
An unresolved continuum source is
also detected in the OVRO data, with a flux of 8.4$\pm$\,1.7 mJy.

\subsection{Position-Velocity Structure}

A position-velocity plot along the bipolar axis (Fig.\,\ref{pv-hst}) 
shows abrupt increases in radial velocity at
offsets of about $\pm$1$''$ from the center. Such structure is typical of
what is
expected when a high-velocity bipolar collimated outflow or jet (hereafter
simply referred to as the jet) interacts
with dense ambient material, producing bow-shocks at the head of the jet
(e.g., see models and Fig.2, right panel, in Lee \& Sahai 2003: hereafter
LS03). The bow-shocks implied by the CO
velocity structure appear to be located in the general vicinity of, and
may be identified with, the bright regions S1, S2 in the W-lobe, and
their presumed optically obscured counterparts in the E-lobe. The P-V
structure has a high degree of symmetry around the center, indicating
that the bipolar outflow and surrounding medium are correspondingly
symmetric.

In addition to the high-velocity bow-shock structures, the P-V diagram
shows a weaker and significantly more spatially-extended component, which
extends out to the tips of the optical bipolar nebula. This component is
composed of relatively low-velocity emission, but the limits of its
velocity structure at the high end cannot be separated from the bow-shock
emission.

There are two possible explanations for the simultaneous presence of the
low- and high-velocity emission structures, with (1) their different
spatial extents seen in the CO J=3--2 P-V plot and (2) correspondence to
the morphological features seen in the optical images. One possibility is
that the extended lobes in I\,22036 are geometrically distinct from the
outflow which produces the high-velocity emission and the bright optical
knots S1,S2. The latter is probably a collimated outflow whose
axis is tilted significantly towards the observer, whereas the extended
lobes are more in the plane of the sky (assuming that the 3-D outflow
velocity of the material in these lobes is not too different from that in
the high-velocity outflow). So in 3-D, I\,22036 may be a quadrupolar
nebula, with the near-side lobe representing the S1,S2 outflow seen in
projection in front of the extended lobe. A second possibility is that
the extended lobes have a similar inclination as the material in the fast
bipolar outflow, and the emission with low (projected) velocity in the
P-V plot comes from an outflow which is intrinsically slower than the
former.

A position-velocity plot along the minor axis of the I\,22036 shows no
significant gradient (Fig.\,\ref{pv-minor}).

\section{The Post-AGB Bipolar Outflow}

\subsection{Kinematics and Expansion Age}

We have evidence of fast outflows in I\,22036 from three independent
probes. Two of these are the CO and H$\alpha$ data in this paper, and a
third is the VLA mapping of OH maser emission reported in SZSCM3 and
Zijlstra et al. (2001). The OH maser features cover a total velocity
spread of about 76\,\kms, and are roughly aligned along the nebular axis,
with a spatial extent comparable to the distance between the
CO-identified bow-shock locations. The P-V signature of the OH is thus
similar to that of CO except at the bow-shock locations, where there is
an absence of OH emission extending to the large outflow velocities seen
in CO. Either OH is destroyed, or the appropriate conditions for exciting
observable maser emission are not met, in and near the bow-shock region. 

Both the OH and CO emission presumably come from molecular gas in the
dense walls of an elongated lobe presumably resulting from the interaction
of a jet
with an ambient circumstellar envelope as shown by LS03. In contrast to
OH and CO emission, the H$\alpha$ absorption probes atomic gas. Since the
H$\alpha$ absorption feature is not spatially resolved in our data, it is
difficult to definitively establish its relationship to the fast
molecular outflow. In the jet-envelope interaction model, the atomic gas
may be located in the interior of the lobes, representing the pristine,
unshocked material of the underlying jet, and/or the interface between
the latter and the lobe walls. We would thus expect the observed velocity
of this atomic material to be generally larger than that of the dense
material everywhere in the lobe. But the blue-shift of the H$\alpha$
absorption feature from the systemic velocity implies a projected outflow
speed of about 100\,\kms, which is smaller than that seen in the CO at
the bow-shocks. This discrepancy can be explained in the context of the
LS03 models, if the H$\alpha$ absorption feature is produced in interface
material
(but not ``pristine", unshocked jet material) at low latitudes in the
outflow lobe,
since the velocity of the interface material near the lobe walls is found
to decrease with latitude (where the outflow axis represents the polar
axis) whereas that of the unshocked jet is constant. 

The LS03 modelling shows that, in the momentum-driven case for a
jet-envelope interaction, the velocity field of much of the dense
material in the resulting elongated lobe is predominantly radial; the age
of the outflow is well approximated by dividing the radial distance of
the bow-shock at the lobe-head by the radial velocity of the material
there. If we take 220\,\kms~and 1$''$, respectively, as the projected
radial velocity and radial distance of the material at the bow shock,
based on Fig.\,\ref{pv-hst}, and if we assume an intermediate
inclination angle ($i=30\arcdeg$) of the outflow axis to the sky-plane,
then the fast outflow age is only 25\,(tan $i$/0.58)\,yr.\footnote{The
much longer estimate of the age (250 yr for an inclination of
$30\arcdeg$) given by SZSCM03 was due to (i) the use of a factor 5 smaller
outflow speed based on the OH data, and (ii) it applied to knot $K_1$
which is located a radius about twice as large as that of the bow shocks}

\subsection{High-Velocity Outflow: Mass and Scalar Momentum}

We now estimate the mass, $M_{bip}$, and scalar momentum, $P_{sc}$
in the high-velocity bipolar outflow from the
integrated J=3--2 line profile shown in Fig.\,\ref{co32spec}a. In this
analysis, we have used the J=3--2 emission flux over a velocity range
extending from 424 (-505)\,\kms~to 15 (-15)\,\kms~from the systemic
velocity for the red (blue) component of the outflow. We have thus
excluded the emission in a velocity range $\pm$15\,\kms~centered on the
systemic velocity, which, as suggested by the $^{13}$CO J=1--0 and the SO
line
profiles, might come from a kinematic component distinct from the bipolar
outflow. We find values of the velocity-integrated J=3--2 emission flux
over the blue and red components of the outflow to be 168.3 and 192.5
Jy\,\kms, repectively.

\subsubsection{Excitation Temperatures and Optical Depths}

We need to determine the excitation temperature and evaluate any optical
depth effects in order to compute the mass of emitting material. The ratio
of the integrated fluxes of the the CO J=1--0 and 3--2 lines, $R13$, over
the high-velocity range covered by both the OVRO and SMA observations (i.e.
an interval of 215\,\kms~centered at the systemic velocity, but excluding
the central $\pm$15\,\kms) is 0.081, lower than the value in the central
$\pm$15\,\kms, which is 0.12. For optically thin emission in both J=1--0
and 2--1, $R13=0.081$ implies an excitation temperature, $T_{ex}$=14.7\,K
and an emitting mass of 0.022\,\ms. However, if either or both of lines are
not optically thin, then the inferred temperature and mass will be
different.

In order to accurately estimate optical depths as a function of velocity, 
a detailed spatio-kinematic model is needed, however, we do not attempt
that here since the emission is not sufficiently well-resolved in our
data, and so the detailed geometrical structure of the emission is poorly
known. Nevertheless, rough estimates of the optical depths can be made by
considering the sizes, total fluxes and total velocity dispersion of the
low- and high-velocity components. We first deconvolve the J=3--2 line
profile into a central gaussian with a width (FWHM) of 30\,\kms~which
represents the low-velocity component, and two pairs of high-velocity
gaussian components (HV1r,HV1b and HV2r,HV2b) whose sum represents most of
the emission from the high-velocity component. Assuming that the centers of
the near (far)
pair, HV1r and HV1b (HV2r and HV2b), are symmetrically offset from the
systemic velocity by $\pm$V1 ($\pm$V2), and their width is W1 (W2), we find
the 
values of V1, V2, W1, W2, as well as the peak intensities of the gaussians
which provide the best fit to the J=3--2 line. We find W1$\sim$92\,\kms~and
W2$\sim$220\,\kms. The total flux in the low-velocity component is
100\,Jy--\kms, whereas HV1 and HV2 have about 191 and 170\,Jy--\kms~each.

The angular extent of the high--velocity component is estimated to be about 
1\farcs2, i.e., 3.6$\times10^{16}$ cm from the J=3--2 blue and red wing
emission maps (Fig.\,\ref{co32wings-hst}). In this estimation, we have
used geometric means of the major and minor axis of both the emission
regions and the beam to obtain deconvolved sizes. Although the detailed
spatio-kinematic structure of the emitting regions is not known, if we
assume that the large velocity gradient approximation is valid, we can
obtain rough estimates of the optical depths. We have simulated a large
velocity gradient by assuming a spherically expanding cloud with constant
expansion velocity equal to half the FWHM of the model components
determined above. Such a cloud has significant velocity gradients along
most lines-of-sight (except those which pass very close to the center).
For each of the four high-velocity gaussian components above, we have
computed the optical depths (and thus 
the masses) and excitation temperatures needed
to fit the R13 ratio and produce half the CO J=3--2 flux of each
component (since the fractional flux within the FWHM points of the
2-dimensional gaussians representing each of the HV1 and HV2 components
is 0.5). These masses are then scaled up by a factor of 2 in order to
obtain the mass corresponding to the total flux in each of the HV
components.  

We have assumed a fractional $f_{CO}$=CO/H$_2$ abundance of
$2\times10^{-4}$ (SZSCM03). For each of the red and blue parts of the HV1
(HV2) component, we find $T_{ex}$=18 (15.5)\,K, line-of-sight J=3--2
optical depth at a radius half the radius of the model cloud,
$\tau$$_{32}$ of 1.1 (0.35), and a mass of $4.4\times10^{-3}$
($3.1\times10^{-3}$)\,\ms. Adding together the HV1 and HV2 masses from the
red and blue parts, and doubling the resulting value to account for the
mass contributing to the flux outside the half-power points of the gaussian
fits to these ccomponents, we get a total emitting mass in the
high-velocity outflow of $M_{bip}=$0.03\,\ms. 

In order to calculate the scalar momentum, we have used the formulation as
described in Bujarrabal et al. (2001), assuming optically-thin
emission, and a single temperature for the high-velocity outflow, which we
take to be the emission-weighted average (17\,K) of the HV1 and HV2
components above. We find $P_{sc}=1.1\times10^{39}$ g cm s$^{-1}$ for an
intermediate inclination angle ($i=30\arcdeg$) of the
nebular axis to the sky-plane. This outflow cannot be driven by radiation
pressure: the dynamical time-scale of the lobes, for $i=30\arcdeg$, is
only $\sim$25\,yr, much smaller than the time required by radiation
pressure to accelerate the observed bipolar outflow to its current speed,
$t_{rad} \gtrsim P_{sc}/(L/c)\sim 10^5$ yr, given I\,22036's luminosity
of 2300\ls. The value of $t_{rad}$ is a lower limit because only a
fraction of the total luminosity, namely that directed into the solid
angle representing the opening angle of the bipolar outflow, may be
utilised for driving the latter.

\section{Molecular Mass versus Dust Shell Mass}
\label{totmass}
\begin{figure}
\epsscale{0.78}
\plotone{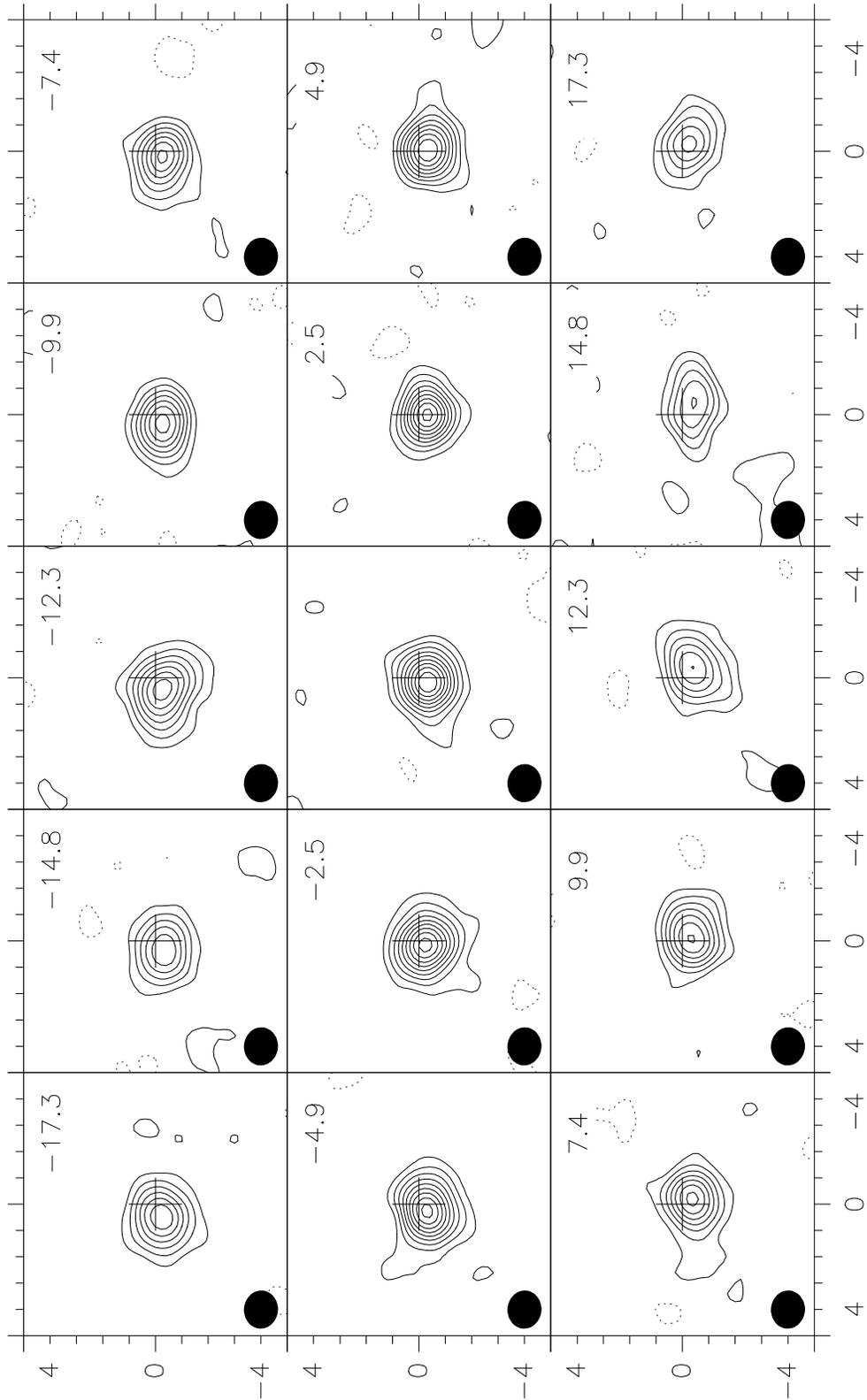}
\caption{As in Fig.\,\ref{chanmap32}, but with the total velocity coverage
extending over the central $\pm$15\kms~range, with each panel showing
emission in a velocity bin of width 2.47\,\kms. Contour levels are 
-10,10,20,...90\% of the peak flux in the central
channel which is 2.2 Jy/beam}
\label{chancen32}
\end{figure}

\begin{figure}
\plotone{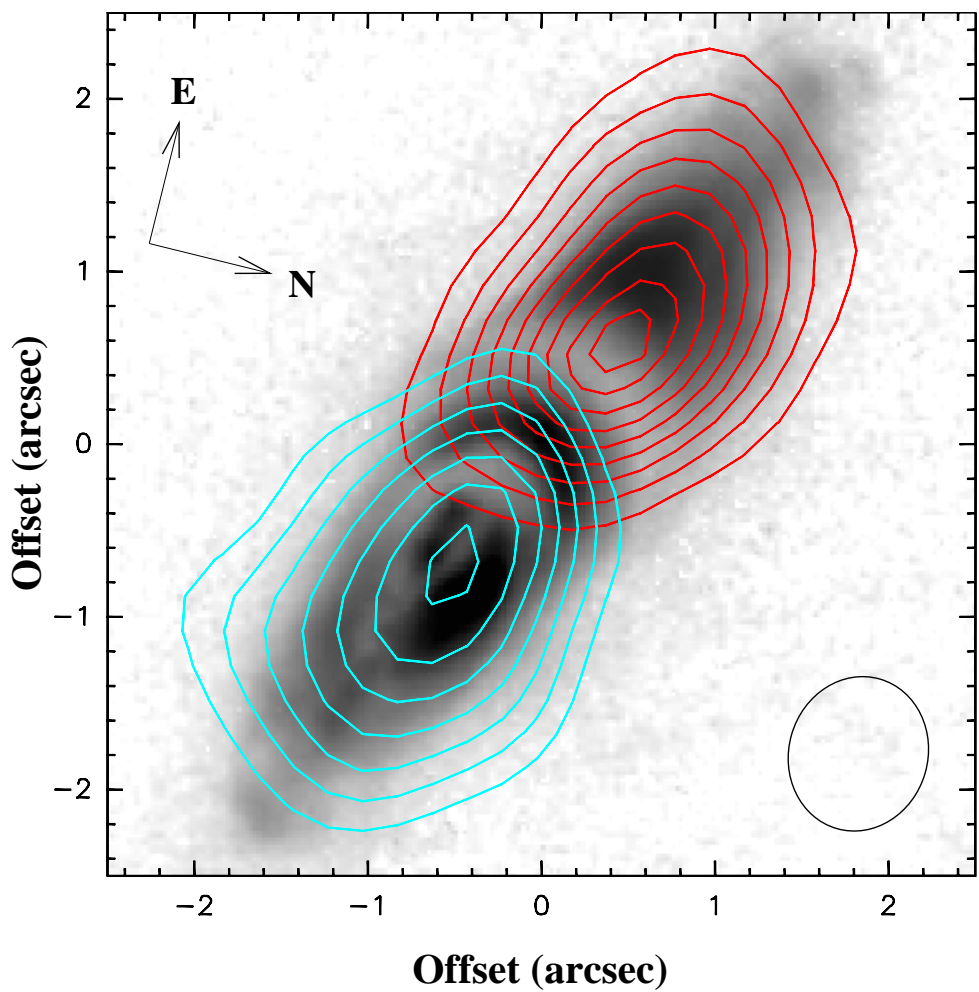}
\caption{Map of the blue and red wing CO J=3--2 line emission from
IRAS\,22036+5306, overlaid on a 0.6\,\micron~HST image. The 
velocity ranges used to define the blue and red wings in this plot are
offset by -245
to -61\,\kms, and 61\,\kms~to 245\,\kms~from the systemic velocity of 
-45\,\kms. The peak surface brightness in the red (blue) lobe is 4.10 (3.37) 
Jy--\kms/beam. Countours are 0.82, 1.23, 1.64... 3.28
Jy--\kms/beam in the blue lobe, (i.e. with a step of 0.41 Jy-\kms/beam),
and 0.82, 1.23, 1.64, ... 3.69, 3.95 Jy--\kms/beam in the red lobe.  The
size and orientation of the clean beam ($1\farcs12\times1\farcs00$, PA
88$\arcdeg$.6) are shown in the lower-right corner}
\label{co32wings-hst}
\end{figure}

\begin{figure}
\plotone{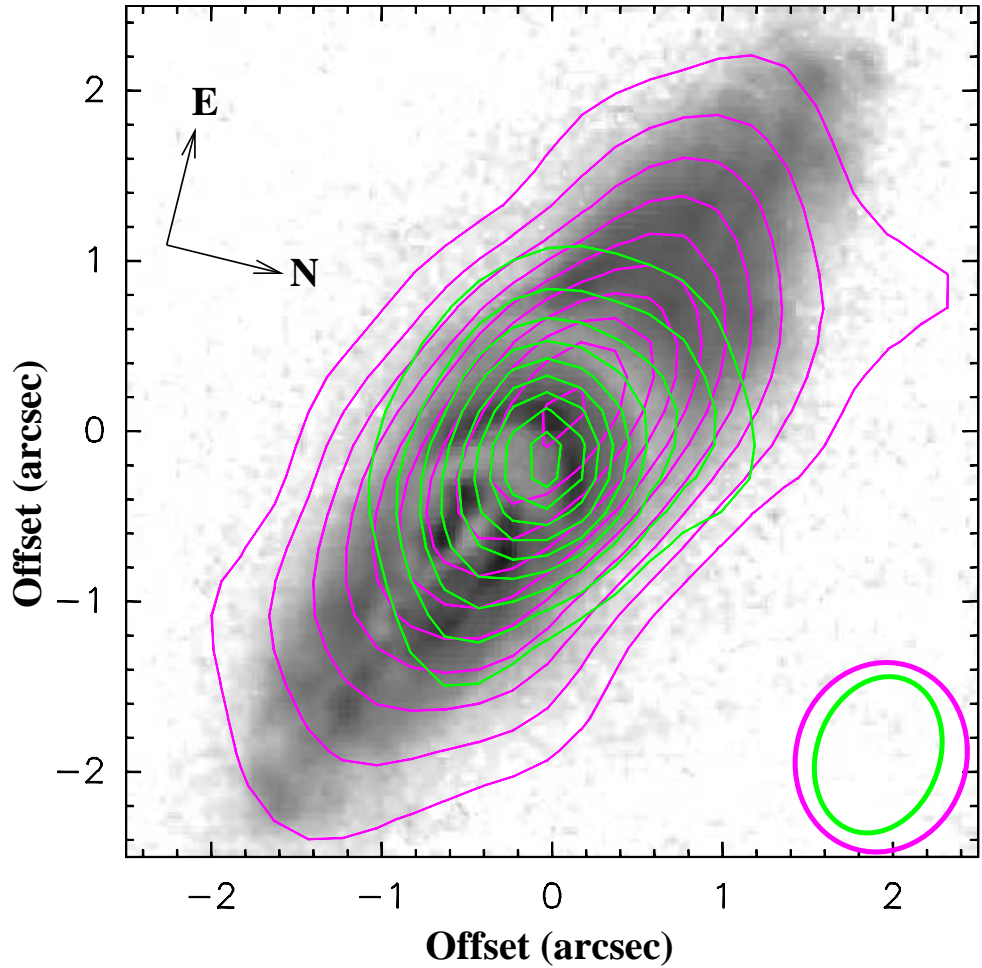}
\caption{Maps of the velocity-integrated CO J=3--2 line emission [magenta
contours; levels are 15 to 95\% of the peak emission (13.3 Jy--km
s$^{-1}$/beam) in steps of 10\%] and 0.88\,mm continuum emission [green
contours; levels are 15 to 95\% of the peak emission (0.17 Jy/beam) in
steps of 10\%] from IRAS\,22036+5306, obtained with the SMA, overlaid on
a 0.6\,\micron~HST image (processed to enhance sharp features, as in
SCZCM3). The 1$\sigma$ noise in the continuum map is 4.5 mJy/beam. The sizes
and orientations of the clean beams for the line maps
($1\farcs12\times1\farcs00$, PA 88$\arcdeg$.6) and continuum maps
($0\farcs94\times0\farcs72$, PA 83$\arcdeg$.9) are shown in the lower-right
corner
}
\label{co32-hstbw}
\end{figure}

\begin{figure}
\plotone{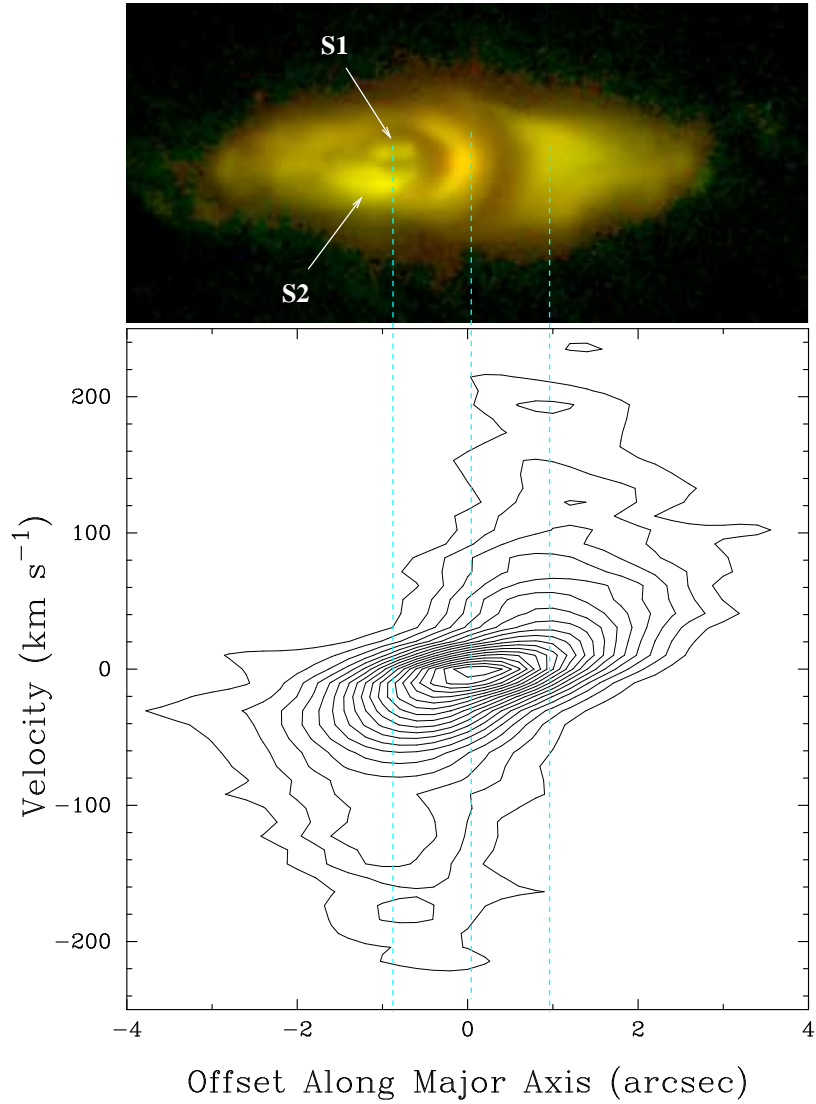}
\caption{Position-velocity plot of the CO J=3--2 emission along
the major axis of the nebula. Contour levels are 5 to 95\% of the peak in
steps of 5\%. A color HST image of the nebula (with the
0.6\,\micron~image in green and the 0.8\,\micron~image in red, as in
SZSCM03)
is shown, with its major axis aligned along the spatial axis in the P-V
plot, for comparison}
\label{pv-hst}
\end{figure}

\begin{figure}
\plotone{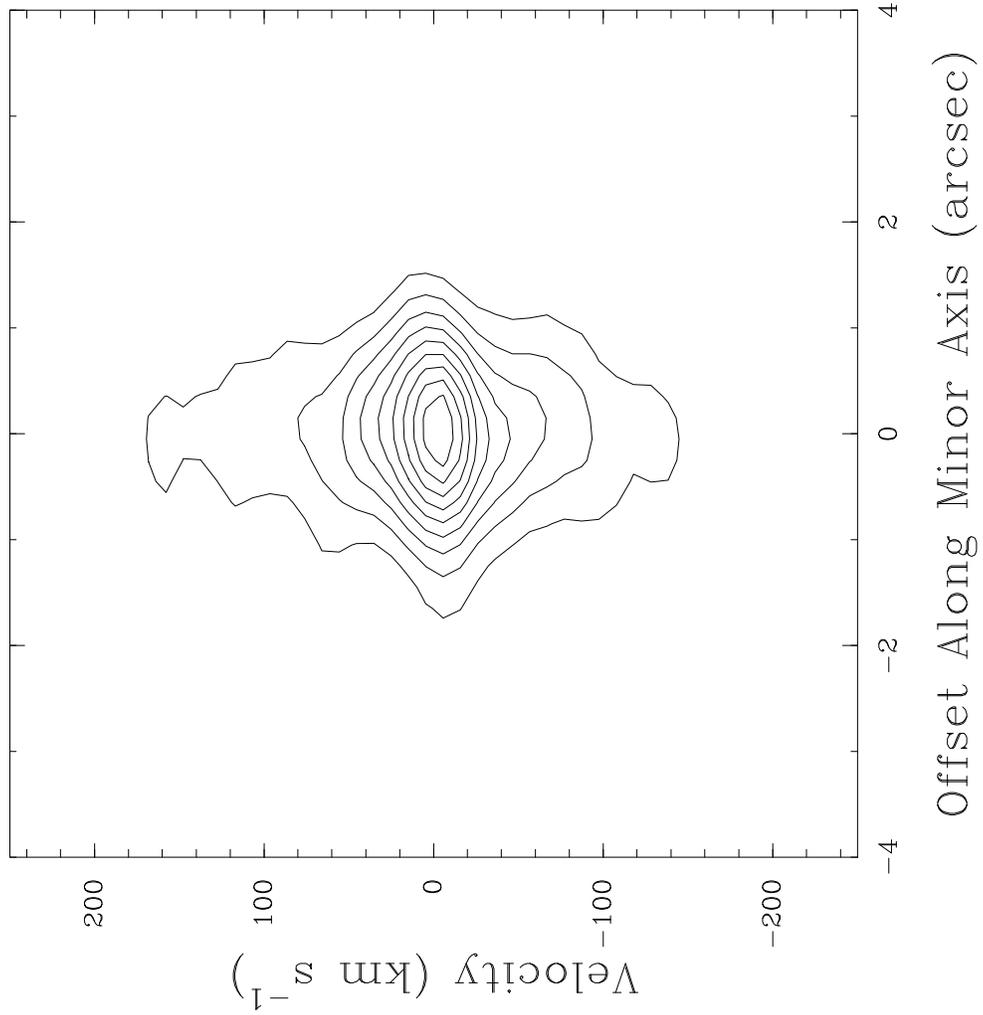}
\caption{Position-velocity plot of the CO J=3--2 emission along
the minor axis of the nebula. Contour levels are 10,20...90\% of the peak in
steps of 10\%}
\label{pv-minor}
\end{figure}

The size of the low--velocity component is estimated to be about 1\farcs3 
(4$\times10^{16}$), from the J=3--2 emission map in Fig.\,\ref{chancen32}.
The flux and angular size of this component are comparable to that in each
of the red and blue components in the HV1 and HV2 features, but it is
spread over a much smaller velocity range. Hence it is expected to be
significantly more optically thick than the HV1 and HV2 components,
making the mass determination more uncertain. Fitting this component as
above, we can find reasonable fits with $T_{ex}\sim26-33$\,K,
$\tau$$_{32}$ about 7.5--10, and a mass of about $0.042-0.035$\,\ms. The
corresponding optical depth of the J=1--0 line, $\tau$$_{10}$ is
$\sim$2.6-2.2 in these models, hence any significant reduction in the
total mass of the shell reduces $\tau$$_{10}$ below unity in large parts
of the emitting region, resulting in a value of R13 which is too low
compared to the observed value of 0.12. For example, if we reduce
$\tau$$_{10}$ by a factor 2, we get R13=0.08, inconsistent with our data.
Thus the minimum mass of the low--velocity component is about 0.035\,\ms.

Our data support the relatively high values of $T_{ex}$ for the
low--velocity component derived from our modelling. The relatively high
surface brightness observed at the systemic velocity (2.2 Jy/beam,
Fig.\,\ref{chancen32}) implies a brightness temperature of
about 12\,K. Accounting for a beam-dilution factor of 2 (since the source
size is about the beam size), the actual source brightness temperature is 
$T_b$=24\,K, and the corresponding excitation temperature required for 
optically thick emission is 32\,K.

Thus the total molecular mass is $M_{tot}=$0.065\,\ms. 
This is much lower than the value ($\sim$4.7\ms)
estimated from the dust mass ($\sim$0.023\ms) inferred from a model of
I\,22036's near to far-infrared fluxes (as observed by ISO, MSX and IRAS)
(SZSCM03), assuming a gas-to-dust ratio, $\delta$=200. In this model, cool
(67--35\,K) dust resides in a large spherical shell (with inner and outer
radii of $1.4\times 10^{17}$ and $5.3\times 10^{17}$ cm) of radial
optical depth $A_V\sim1.5$, and hot (1000-350\,K) dust resides in a
very optically-thick ($A_V\sim120$), compact ($(3-4.6)\times 10^{14}$ cm)
central torus. The mass of the cool shell is 4.7\ms, and
the (very poorly constrained) mass of the compact central torus is 
$6\times10^{-4}$\ms, assuming a 100$\mu$m dust emissivity of 34
cm$^2$\,g$^{-1}$. An additional warm (145--82\,K) shell (with inner and
outer radii of $1.3\times 10^{16}$ and $5.7\times 10^{16}$ cm) with
relatively little mass $\sim$0.005\ms~is added to the first two
components to obtain a good fit to the spectral energy distribution (SED)
in the $\sim$20-40\,\micron~range.

The size of the CO J=3--2 emission source along its long axis is $\sim
5''$, which is significantly smaller than the central cavity (of size
$\sim 9''$) in the model cool dust shell. Thus the observed CO J=3--2
emission does not probe the material in the cool shell, which explains
the discrepancy between the mass estimate based on the CO data
($\sim$0.032\ms) and that derived from the dust emission (4.7\ms). The
size of the dust shell is large (inner and outer radii of 4\farcs6 and
17\farcs6) compared to the typical size of structures which would not be
largely
resolved out by the interferometer, which is $\sim9''$ given our minimum
UV spacing of 10 klamba, and good UV coverage for spacings greater or
equal
to about 23 klambda. Hence it is likely that a significant fraction of
the CO J=3-2 flux from this shell is resolved out by our interferometric
observations. Furthermore, astrophysical mechanisms for the lack of
observable CO J=3--2 emission from the dust shell are also plausible,
e.g. (a) photodissociation of CO due to the interstellar ultraviolet 
radiation, and (b) insufficient excitation due to low gas kinetic
temperatures. We think it is unlikely that photodissociation is
important, since the photodissociation radius (estimated using Table 3
from Mamon et al. 1988) given the high
mass-loss rate inferred for the shell ($5.5 \times 10^{-4}$\my: SZSCM03),
and an assumed expansion velocity of 15\,\kms, is a few\,$\times\,10^{18}$
cm, which is significantly
larger than the outer radius of the dust shell. As in the case of other
PPNs (e.g., IRAS 19024+0044: Sahai et al. 2005, Roberts 22: Sahai et al.
1999), it is plausible that the gas kinetic temperatures in the extended
dust shell of I\,22036 are quite low because of the combined effect of
cooling due to adiabatic expansion and the lack of dust frictional
heating.

\subsection{The Central Submillimeter Continuum Source: Large Grains in
I\,22036's Dusty Torus}
\begin{figure}
\plotone{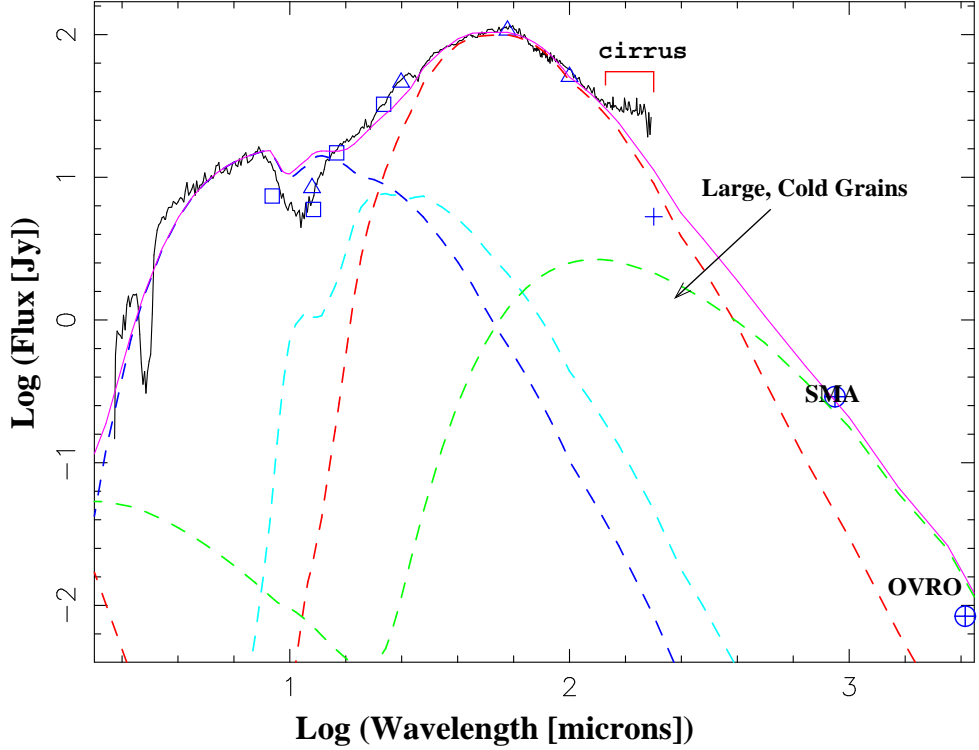}
\caption{Observations [{\it black curve}: ISO spectra, {\it blue symbols}:
photometric data -- SMA 0.88\,mm and OVRO 2.6\,mm: {\it circles
w/crosses}, MSX: {\it squares}, IRAS: {\it triangles}, ISO/PHOT: {\it
cross}] and a model spectrum ({\it magenta curve}) of I\,22036.
Individual components ({\it dashed curves}) of the model are also shown:
cool ({\it red}) \& warm ({\it cyan}) shells, hot inner disk ({\it blue})
and a component with large, cold grains ({\it green}), probably
associated with I\,22036's dusty waist. The flattened region at the
long-wavelength end of the ISO spectrum is most likely due to
contamination from interstellar cirrus emission.}
\label{iso-submm}
\end{figure}

The 0.88\,mm and
2.6\,mm fluxes reported here are significantly in excess of the SED
resulting from the dust model of SZSCM03. This
can be seen in in Fig.\,\ref{iso-submm}, which shows a modified version
of the SZSCM03 model (discussed below) -- the SMA and OVRO data points lie
significantly above the dashed red curve representing emission from the
cool shell, which is the dominant component at long wavelengths in the
SZSCM03 model. Since the observed 0.88\,mm continuum source is unresolved,
its size is less than $0\farcs9\times0\farcs7$ (the FWHM of the SMA clean
beam). Thus the 0.88\,mm flux in the SZSCM03 model, within this small
aperture, would be even lower than shown by the dashed red curve in
Fig.\,\ref{iso-submm}. Since the spectral index of the continuum from
0.88\,mm to 2.6\,mm, $\alpha$ (defined as $F_{\nu} \sim \nu^{\alpha}$)
is 3.3, too steep to be a result of thermal brehmstrahhlung, the
submillimeter and millimeter-wave fluxes in I\,22036 are most likely the
result of thermal emission from a population of quite large, cold grains.
We believe that the grains must be large because the  
opacity spectral index 
$\beta=d\,ln\,\kappa/d\,ln\,\nu$ in the submillimeter-millimeter wavelength region 
from our data\footnote{derived assuming 
optically-thin emission in the Rayleigh-Jeans limit, for which $\alpha=2+\beta$}
is 1.3, which is significantly smaller than the 
value expected for standard small grains ($\beta\approx$2). Such small 
values of $\beta$ are consistent with those obtained in theoretical models with
grain sizes  
(in a power-law size distribution) extending to the millimeter range (Draine
2006).  
Given the compact nature of the 0.88\,mm continuum source, it is quite
likely that these large grains are physically located within the dusty
waist seen in the HST images. Note that the bulk of the grains producing
the SED of I\,22036 in the $\sim$few-200\,\micron~range are probably small
($\sim$0.1\,\micron) as has been found for other PPNs from the very large
fractional polarisations typically observed towards these objects at
optical and near-infrared wavelengths (e.g. Sahai et al. 1998a, 1999,
Oppenheimer et al. 2005).

In order to provide rough constraints on the physical parameters of the
large-grain dust source, we have used DUSTY (as in SZSCM03) to compute the
emission from a dust shell with large grains, varying its major inputs --
the temperature at the shell inner radius (the ratio of the outer to
inner radius is arbitrarily assumed to be 2), the optical depth at
100\,\micron, $\tau(100)$, the grain temperature at the inner radius
$T_d(in)$, and the grain size, $a$ -- to fit the observed SMA and OVRO
continuum fluxes. We find that dust grains with $a\gtrsim$1\,mm~and
$T_d(in) \lesssim$ 50\,K are needed for producing the observed fluxes at
0.88 and 2.6\,mm. The shell has an optical depth $\tau(100)=0.1$. 
Models with smaller grains (e.g. $a\lesssim$
500\,\micron) which fit the 0.88 and 2.6\,mm fluxes produce too much flux in
the $\sim$100-200\,\micron~range. In order for the
large-grain shell to be smaller than or equal to the observed size of the
0.88\,mm continuum source, i.e. $ < 2.5\times 10^{16}$\,cm, the
source luminosity corresponding to the required irradiating intensity must
be $\lesssim $350\ls, i.e., $\sim$15\%
of the total source luminosity. This result requires that the large
grains be confined to a region where they can be substantially shielded
from the full radiation of the central star, supporting our conjecture
that they are located within the toroidal waist seen in the HST images of
I\,22036, since this region is partially shielded from the full stellar
radiation field by the compact, optically-thick central dust component
inferred by SZSCM03. From our model, we estimate the mass in dust of this
large grain component to be $\sim$0.04\ms, taking the dust mass
extinction coefficient, $\kappa$, at $\lambda << a$ (say at 100\,\micron)
to be 2.5 cm$^2$g$^{-1}$, derived from 
$\kappa=\sigma (4\pi a^3 \rho_d/3)^{-1}$, with 
$\sigma=\pi a^2$, a=1\,mm and $\rho _d=3$ g\,cm$^{-3}$.

Our dust model is based on commonly-used values for the wavelength-dependent
complex refractive index of silicate dust particles, however uncertainties in the
dust mass absorption coefficient may allow for a reduction of the mass estimate
by as much as a factor
of 2. Thus it is clear that there is a
very substantial mass of such grains in the dusty waist of I\,22036. If
we assume a gas/dust ratio of 200 in the waist, as in the cool dust
shell, the total mass in the toroid becomes 4--8\ms, a disturbingly large
value, especially when combined with the 4.7\,\ms~of material in the cool
dust shell. We therefore conclude that the gas content in the waist must
be significantly depleted, and the gas-to-dust ratio consequently much
lower than the typical value, maybe by as much as an order of magnitude.

\section{The $^{13}$C to $^{12}$C isotope ratio}
We derive the $^{13}$C to $^{12}$C abundance ratio, $f_{13/12}$, in
I\,22036 from the ratio of the J=1--0 $^{13}$CO and $^{12}$CO fluxes. The
difference in the shapes of the $^{12}$CO and $^{13}$CO lines shown in
Fig.\,\ref{co32spec} is consistent with a uniform value of $f_{13/12}$,
but larger optical depths in the central component compared to the
high-velocity component. As indicated by the prominent core emission seen
in the $^{13}$CO but not the $^{12}$CO line profile, the ratio of the
$^{13}$CO to $^{12}$CO flux in the core (0.37) is much higher than that
in the high-velocity wings (0.2). This difference in the flux ratio
between the low-velocity and high-velocity outflows is due to the
difference in their optical depths: $\tau$$_{10}$ is about 2 in the
central component but 0.4 in the HV1 component (see \S\,\ref{totmass}). 
We have therefore computed $f_{13/12}$ from the flux ratio in the more 
optically-thin part of the lines -- the $^{13}$CO/$^{12}$CO flux ratio of
0.2 gives $f_{13/12}=0.16$. We find that the same value of
$f_{13/12}$ reproduces the observed $^{13}$CO to $^{12}$CO flux ratio in
the core if $\tau$$_{10}$=2.2 and $T_{ex}=30$\,K, supporting our modelling
of the $^{12}$CO emission from the low-velocity outflow, 

Thus the $^{13}$C to
$^{12}$C abundance ratio in I\,22036 is close to the maximum value
achieved in equilibrium CNO-nucleosynthesis (0.33), and provides a strong
test of models of stellar nucleosynthesis. For comparison, other
evolved objects which show high values of $f_{13/12}$ (i.e., in excess of
0.1) are (i) the circumstellar envelopes of some oxygen-rich AGB or
post-AGB objects such as OH231.8+4.2 (S\'anchez Contreras et al. 2000),
Frosty Leo (Sahai et al. 2000), 
U Equ (Geballe et al. 2005), (ii) the rare class of J-type carbon stars
(e.g., Y CVn or T Lyr: Lambert et al 1986), and (iii) selected planetary
nebulae (e.g., M\,1-16: Sahai et al. 1994). In general, there is increasing
observational evidence that s large fraction of evolved stars have
$^{12}$C/$^{13}$C abundance ratios lower than about 15, significantly lower
than those predicted by standard models. For example, from millimeter-wave
CO studies, Palla et al. (2000) and Balser, McMullen \& Wilson (2002) have
estimated the $f_{13/12}$ ratio in modest PN samples (16 \& 9 objects,
respectively) and find it lies typically in the range $\sim$0.04--0.1. 

There are two processes which can enhance the $f_{13/12}$ ratio above the
values predicted by the standard model. The first, cold-bottom-processing
(CBP), is a deep mixing mechanism 
believed to operate during the RGB phase of stars with mass less than about
2\ms (Wasserburg, Boothroyd \& Sackmann 1995), and the degree of enhancement
decreases with increasing stellar mass. In a 1\ms~star, CBP can
produce a value of $f_{13/12}$ as high as 0.2 during the RGB phase, but
this ratio is lowered to 0.025--0.05 during the AGB phase due to the
addition of fresh $^{12}$C. Since the progenitor central star of I\,22036
is significantly more massive than 2\ms~considering the high mass of its
dust shell, CBP cannot account for its high $f_{13/12}$ ratio.

The second mechanism, hot-bottom-burning (HBB), is believed to operate in
AGB stars with initial masses above about 3.5--4\ms. Such stars develop
deep convective envelopes with very high temperatures at the base causing
the operation of the CN cycle (i.e., $^{12}$C $\rightarrow$  $^{13}$C  
$\rightarrow$ $^{14}$N). In this case, $f_{13/12}$ will be near the
equilibrium ratio of 0.33, and the surface $f_{13/12}$ is expected to
asymptotically approach this limit. HBB is thus the likely mechanism for
producing the high $f_{13/12}$ ratio in I\,22036. Rubin et al. (2004)
reach a similar conclusion for the PN\,2440, where they find $f_{13/12}$
$\sim$0.2 from analysis of CIII lines in IUE data.

\section{DISCUSSION}
\subsection{The shaping of bipolar Planetary Nebulae}
The dynamical evolution and shaping of PPNs is believed to result from
the shock interaction between a fast, collimated post-AGB wind and the
slowly expanding, dense wind ejected during the previous AGB phase. The
bipolar post-AGB wind impinges on, and thus accelerates and heats, two
localized regions of the (presumably spherical) AGB wind. A significant
amount of momentum is believed to be transferred to the rest of the AGB
shell by means of bow-like shocks. Although morphological
identifications of bow-shocks have been made in PPNs, their
characteristic spatio-kinematic structure has seldom been identified.
Thus, the bow-shock structures inferred kinematically from our CO J=3--2
imaging of I\,22036 provide support for, and are an important input for
quantitative modelling of the physics of, the interaction of collimated
fast winds with AGB envelopes as a production mechanism for bipolar
PPNs (see, e.g., LS03).

We have shown that the bipolar molecular outflow in I\,22036 cannot be
driven by radiation pressure. Bujarrabal et al. (2001) infer the
presence of fast molecular outflows in a large sample of PPNs, from the
extended wings seen in their CO line profiles, and conclude that these
outflows cannot be driven by radiation pressure. Only for a small fraction 
of objects in their sample (e.g., CRL2688 or the Egg Nebula: Cox et al.
2001, IRAS\,09371+1212 or the Frosty Leo Nebula: Castro-Carrizo et al.
2005, CRL\,618: S\'anchez Contreras et al. 2004, OH231.8+4.2: Alcolea et
al. 2001) has
interferometric mapping been carried out, directly revealing the
structure of the fast outflow. In all such well-mapped PPNs, the fast
outflow or each of its components (if it is comprised of multiple
components) appears to be collimated and directed along well-defined
axes. Our results show that I\,22036 fits well in this picture of the
fast molecular outflows in PPNs. Given the increasing direct evidence
for the presence of jet-like outflows in PPNs (e.g., W43A: Imai et al. 2002;
IRAS\,16342-3814: Sahai et al. 2005, IRAS\,19134+2131: Imai et al. 2004),
we believe that the case for the fast
molecular outflows in I\,22036 and other PPNs being driven by
collimated jets is quite strong. The presence of such jets, operating
in an episodic fashion and changing their directionality, has been
proposed as a general mechanism for explaining the diverse morphologies
and widespread presence of point-symmetry in planetary nebulae (Sahai and
Trauger 1998, Sahai 2004). The launching mechanism for such jets
remains unknown, but given the similarity between the fast outflow
speeds in PPNs and jets in low-mass pre-main sequence stars with
accretion disks, it is plausible that the mechanisms which drive jets
in PPNs and YSOs are very similar. In this scenario, the accretion disk
in a PPN would be produced around a relatively close companion low-mass
star, which accretes material from the dense AGB wind of the primary
(Morris 1987, Mastrodemos \& Morris 1998).

Our CO J=3-2 observations and the dust model of SZSCM03 indicate an
apparent absence of dense cool material in I\,22036 along non-polar
directions, in the volume between the molecular outflow and the inner
radius of the cool dust shell (hereafter, the ``in-between" region). In
the SZSCM03 model, the inferred inner
radius of this dust shell cannot be significantly smaller than the
derived value because I\,22036's SED (which peaks near 60\,\micron)
dictates that the emitting grains have low temperatures, and therefore
be located relatively far from the illuminating source. Even if we
explicitly account for the reduction in the illuminating flux due to
the dusty torus in SZSCM03's model, which covers $\sim$0.4 of the 4$\pi$
solid angle around the central star, the lower limit to the fraction of
the central star's total luminosity which irradiates the cool dust
shell is 0.6. Hence the inner radius of the shell (which scales as the
square-root of the illuminating flux) is at most 23\% smaller than
derived by SZSCM03. Although one may explain the absence of CO emission
using insufficient excitation as an argument, the lack of thermal dust
emission from the in-between region is strong evidence for the lack of
dense material there.

A faint halo has been found in I\,22036 at 0.8\,\micron~in new deep images
taken with the HST/ACS camera (Sahai et al. 2006, in
preparation) which indicates the presence of a rather tenuous
surrounding remnant AGB envelope. But we do not know if the fast
molecular outflow in I\,22036 was produced by a collimated jet
interacting with this tenuous envelope. If the ambient material is of
sufficiently low density, then the swept-up shell can be too faint to
be observable, and one can then see the jet material directly, as
appears to be the case for the young PN, Hen 2-90 (Sahai \& Nyman 2000,
Lee \& Sahai 2004). Only detailed numerical simulations of the jet-AGB
envelope interaction in I\,22036 can tell us if the material in its
fast molecular outflow (a) represents the swept-up shell due to the
interaction of the jet with the tenuous AGB envelope seen in the
near-IR halo, (b) is intrinsic, i.e., belongs to the
underlying jet itself, or (c) represents material associated with the dense
torus that has been 
accelerated by the jet. In model $c$, one can imagine that the jet punched
its way out of a flattened dusty cocoon around the central star as
envisaged for the case of the well-studied PPN, CRL\,2688, Sahai et al.
(1998b) -- cocoon matter entrained by the jet produces the observed bipolar
outflow in CO emission, and the remnant cocoon represents the torus.

While considering the above issues, it is useful to note that the
nebular properties of I\,22036 are very similar to that of the
well-studied object, OH231.8+4.2, although their central stars have 
different spectral types ($\sim$F5 versus M9, respectively). As in
I\,22036, OH231.8 harbors a
massive molecular bipolar outflow oriented along the long axis of the
optical bipolar nebula, and a weak round near-IR halo is seen around
its bipolar outflow, signifying the presence of a tenuous spherical
envelope surrounding this outflow (i.e., characterised by a mass-loss
rate of $\sim 10^{-6}$\my: Alcolea et al. 2001). Furthermore, I\,22036
also has a very high $^{13}$C to $^{12}$C ratio, similar to that in
OH231.8. Thus it appears that both I\,22036 and OH231.8 are very similar
objects and 
are formed by a similar mechanism; the main difference between the two
is that OH231.8+4.2 is a more extreme version -- in it, the
mass of cool material in the bipolar outflow is estimated to be
significantly larger (Alcolea et al. 2001). Both objects show
high-velocity OH maser emission. Since OH231.8 was not observed by ISO,
we do not know its far-IR SED as well as that of I\,22036, however the
existing IRAS data show that the SEDs of both objects are very similar.
For example, in OH231.8 (as in I\,22036), (a) the broad-band IRAS
fluxes from 12 to 100\,\micron~show a peak at 60\,\micron, and (b) the LRS
data shows a broad silicate absorption feature at 10\,\micron~and a
steeply rising spectrum towards longer wavelengths. OH231.8 also shows
a compact continuum component at 1.3\,mm (S\'anchez Contreras et al.
1998), which the authors interpret as emission from grains larger than about 
5\,\micron. 

I\,22036 most likely had a fairly massive progenitor, easily $\gtrsim$4\ms.
The combination of the high circumstellar mass (i.e., in the extended dust
shell and the torus) and the high $^{13}$C/$^{12}$C ratio in this object
provides strong support for the operation of the hot-bottom-burning
mechanism.



\subsection{Dusty Waists and Large Grains}
Compared to the question of forming bipolar and multipolar morphologies
in PPNs, the formation of the dense, dusty waists in these objects
remains a much bigger puzzle. The presence of large grains with a very 
substantial mass associated with
I\,22036's toroidal waist, is a surprising, but not unprecedented
result. In the well-studied PPN, AFGL\,2688, Jura et al. (2000) 
find evidence for a substantial mass ($\sim$0.01\ms) of large-sized
grains. In the Red Rectangle (RR: also the nearest PPN to us at a distance
of 350-700\,pc, bipolar in morphology and harboring a disk in 
rotation around a central binary), Jura et al. (1997) derive a minimum mass
of $5\times10^{-4}$\ms~in large ($a\gtrsim$0.02 cm), cool (50\,K) grains
based on RR's 1.3\,mm continuum flux. Men'shchikov et al. (2002) fit the
full SED and near-infrared images of the RR using a 2-D dust radiative
transfer model, and infer a dust mass of about 0.01\ms~in grains with
$a\,\sim$1\,mm in a compact, central torus. In the quadrupolar PPN
IRAS\,19475+3119, Sarkar \& Sahai (2006)
find, from a comparison of a detail model of its full SED (from optical
wavelengths to 
200$\mu$m) to the observed 0.85\,mm~flux, that there is a substantial
0.85\,mm excess flux which also
requires the presence of large, cool grains. We speculate that the formation
of dusty waists and the presence of large grains in PPNs are due to
intimately-linked physical processes, thus a resolution of the issue of
waist formation may well lead 
lead to a resolution of how the large grains are made or vice versa. A possible 
mechanism is the destruction of 
volatile cometary-debris disks within several hundred to a thousand AU 
(Stern et al. 1990) by 
an intermediate-mass star during its luminous post-main-sequence evolution, 
providing a source of large solid particles which could form a 
dusty torus. However, the total 
mass of large grains which we have found in I\,22036 is far too large 
compared to the mass estimates for our Kuiper Belt ($\sim$0.1$M_{\earth}$, e.g., 
Luu \& Jewitt 2002) 
or the inner parts of the Oort Cloud ($\sim$40$M_{\earth}$, e.g., Dones et al.
2004), 
and argues against such a process, unless of course the Kuiper Belt/Oort Cloud
analogs 
of an intermediate mass star like that in I\,22036 are significantly more massive
than 
those in our Solar system.

\section{Conclusions}

Using the SMA, we have made high angular resolution ($\sim 1{''}$) maps of 
the CO J=3--2 emission in
IRAS\,22036+5306, a bipolar pre-planetary nebula with knotty jets,
discovered previously in one of our HST imaging surveys of candidate
PPNs. These data have been supplemented by lower-resolution CO and
$^{13}$CO J=1-0 observations with OVRO, as well as optical long-slit
echelle spectroscopy. We find 

\noindent (i) The CO J=3--2 spatially integrated line profile covers a
total velocity extent (FWZI) of $\sim$500\,\kms. The CO and $^{13}$CO
J=1--0 also show large velocity widths, although with a smaller spread
due to the lower sensitivity of these data.\\
(ii) The CO J=3--2 emission comes from a massive high-velocity bipolar
outflow aligned with the optical nebulosity. The characteristic 
position-velocity signature of a bow-shock resulting from a high-velocity
bipolar collimated outflow/jet interacting 
with dense ambient material can be seen in the CO J=3--2 data in both the
red and blue-shifted components of the outflow. The P-V
structure has a high degree of symmetry around the center, indicating
that the bipolar outflow and surrounding medium are correspondingly
symmetric. The bow-shocks do not lie at the tips of the optical lobes.\\
(iii) The H$\alpha$ line shows a profile with a central peak, very wide
wings, and an absorption dip in the blue wing of the line, blue-shifted
from the systemic velocity by about 100\,\kms. The absorption feature
indicates the presence of a fast, neutral, outflow inside the lobes.\\
(iv) A central, unresolved, continuum source with a flux of
0.29$\pm$0.04\,Jy (8.4$\pm$1.7\,mJy) at $\lambda$=0.88\,mm (2.6\,mm) is
seen in the SMA (OVRO) data.\\ 
(v) The mass in the high-velocity outflow, i.e., at velocities offset
more than $\pm$\,15\,\kms~from the systemic velocity, is $0.03$\,\ms, and
the
scalar momentum is 1.1$\times10^{39}$ g cm s$^{-1}$. The large value of
the scalar momentum, together
with the small age of the high-velocity outflow, implies that it cannot
be driven by radiation pressure.\\
(vi) The total molecular mass derived from integrating the flux of the CO
J=3--2 line over its full velocity range, is
$\sim$0.065\ms, much lower than the value ($\sim$4.7\ms)
estimated previously from a dust-shell model of
I\,22036's near to far-infrared fluxes, assuming a gas-to-dust ratio,
$\delta$=200. This is probably due to the interferometric J=3--2
observations resolving out the flux from the large shell (inner and outer
radii of $1.4\times 10^{17}$ and $5.3\times 10^{17}$ cm) in which the cool
dust resides. Furthermore, the gas in this shell may not be sufficiently
excited.\\  
(vii) The small size of the continuum source indicates that it is 
associated with the dusty toroidal waist of I\,22036. The submillimeter
and millimeter-wave continuum fluxes imply the presence of a very
substantial mass ($\sim$0.02-0.04\ms) of large (of radius $\gtrsim$
1\,mm) cold grains in this waist.\\
(viii) The $^{13}$C/$^{12}$C ratio in I\,22036 is 0.16, close to the
maximum value value achieved in equilibrium CNO-nucleosynthesis (0.33). The
enhancement is most likely due to the operation of hot-bottom-burning,
which operates in stars with initial masses larger than about 4\ms.

We thank Nick Sterling for pointing out the importance of hot-bottom-burning
in enhancing the $^{13}$C/$^{12}$C ratio, and Noam Soker for reading an
earlier version of this paper. RS is thankful for partial financial support
for this work from a NASA/ ADP and LTSA grant. CSC's work was supported by
the National Science Foundation through Grant No. 9981546 to Owens Valley
Radio Observatory and the Spanish MCyT under project DGES/AYA2003-2785 and
the 2005 ``Ram\'on y Cajal" program.

\clearpage

\begin{table}[]
\caption{IRAS22036+5306 Model Components}
\begin{center}
\begin{tabular}{llll}
\tableline\tableline
Component         & Parameter & Value & Comments \\
\tableline\tableline

Cool,             & R$_{in}$    & 1.4$\times10^{17}$cm & SZSCM03\\
Dense Shell       & R$_{out}$   & 5.3$\times10^{17}$cm & \\
                  & T$_d$(in)  & 67\,K              & \\
                  & T$_d$(out) & 35\,K              & \\
              & density law    & r$^{-2}$  & \\
                  & Mass       & 4.7\ms & \\
                  & \mdot    & 5.5$\times10^{-4}$\my   & \\
                  & grain size &  MRN     & \\
\tableline
Warm Shell        & R$_{in}$   & 1.3$\times10^{16}$cm & SZSCM03\\
                  & R$_{out}$  & 1.9$\times10^{16}$cm & \\
                  & T$_d$(in)  & 145\,K   &  \\
                  & T$_d$(out) & 82\,K    &  \\
                  & Mass       & 5.5$\times10^{-3}$\ms&   \\
                  & grain size &  MRN     & \\
\tableline
Inner Torus/Disk  & R$_{in}$   & 3$\times10^{14}$cm & SZSCM03\\
                  & R$_{out}$  & 4.5$\times10^{14}$cm & \\
                  & T$_d$(in)  & 1000\,K   &  \\
                  & T$_d$(out) & 365\,K    &  \\
                  & Mass       & 6$\times10^{-4}$\ms&   \\
                  & grain size &  MRN     & \\
\tableline
Toroid            & R$_{in}$   & 9$\times10^{15}$cm & this paper\\
                  & R$_{out}$  & 1.8$\times10^{16}$cm & \\
                  & T$_d$(in)  & 50\,K     &    \\
                  & T$_d$(out) & 35\,K     &   \\
                  & Mass(dust) & 0.02-0.04\ms   &  \\
                  & grain size & $a\gtrsim$1\,mm  & \\
\tableline		
Fast Mol. Outflow & Speed    & $\sim$220\,\kms & this paper\\ 
                  & Age      &  25\,yr        &  \\
                  & Mass     & 0.030\ms &  \\
	          & Momentum & 9.2$\times10^{38}$ g cm s$^{-1}$& \\
                  & $t_{rad}$  & $\gtrsim10^5$yr & \\
\tableline		
Slow Mol. Outflow & Speed    & $\sim$15\,\kms & this paper\\
                  & Mass     & 0.035\ms &  \\
\tableline
Fast Atomic Outflow & Speed    & $\sim$100\,\kms & this paper\\
\tableline
\end{tabular}
\end{center}
\label{tabphot}
\end{table}

\clearpage

%

\end{document}